\documentclass[%
nofootinbib,
 amsmath,amssymb,
 aps,
twocolumn,
]{revtex4}

\usepackage{graphicx}
\usepackage{dcolumn}
\usepackage{bm}

\usepackage{braket}
\usepackage{xcolor}
\usepackage{cancel}
\usepackage{titlesec}
\usepackage{etoolbox}

\renewcommand{\vec}{\bm}

\newcommand{\be}{\begin{equation}}
\newcommand{\ee}{\end{equation}}
\newcommand{\Omegazero}{\Omega^{(0)}}
\newcommand{\dd}{\mathrm{d}}
\newcommand{\grad}{\vec{\nabla}}

\begin{document}

\preprint{APS/123-QED}

\title{Fluctuation relations for systems in constant magnetic field}

\author{Alessandro Coretti}
\affiliation{ORCID: 0000-0002-7131-3210}
\affiliation{%
 Department of Mathematical Sciences, Politecnico di Torino, Corso Duca degli Abruzzi 24, I-10129 Torino, Italy
}%
\affiliation{
Centre Europ\'een de Calcul Atomique et Mol\'eculaire (CECAM), \'Ecole Polytechnique F\'ed\'erale de Lausanne, Batochime, Avenue Forel 2, 1015 Lausanne, Switzerland
}%


\author{Lamberto Rondoni}
\affiliation{ORCID: 0000-0002-4223-6279}
\affiliation{%
Department of Mathematical Sciences, Politecnico di Torino, Corso Duca degli Abruzzi 24, I-10129 Torino, Italy%
}%
\affiliation{
Istituto Nazionale di Fisica Nucleare, Sezione di Torino, Via P. Giura 1, I-10125 Torino, Italy
}%

\author{Sara Bonella}
\email{sara.bonella@epfl.ch}
\affiliation{ORCID: 0000-0003-4131-2513 \\
Centre Europ\'een de Calcul Atomique et Mol\'eculaire (CECAM), \'Ecole Polytechnique F\'ed\'erale de Lausanne, Batochime, Avenue Forel 2, 1015 Lausanne, Switzerland
}%


\date{\today}

\begin{abstract}
The validity of the Fluctuation Relations (FR) for systems in a constant magnetic field is investigated. Recently introduced time-reversal symmetries that hold in presence of static electric and magnetic fields and of deterministic thermostats are used to prove the transient FR without invoking, as commonly done, inversion of the magnetic field. Steady-state FR are also derived, under the t-mixing condition. These results extend the predictive power of important statistical mechanics relations. We illustrate this via the non-linear response for the cumulants of the dissipation, showing how the new FR enable to determine analytically null cumulants also for systems in a single magnetic field.


\end{abstract}

\maketitle

\section{Introduction}
Statistical mechanics has traditionally investigated macroscopic systems at or near 
thermodynamic equilibrium, where fluctuations of observables are negligible compared to their mean value.
More recently, however, nano- and bio-sciences have called attention to mesoscopic scales, in which fluctuations are considerably more relevant~\cite{stassi:2017,bonaldi:2009} and the notion of thermodynamic equilibrium  problematic.
Consequently, theories of fluctuations and of far-from-equilibrium response, have become a major chapter of contemporary statistical mechanics. 
In particular, a fruitful line of research on non-equilibrium fluctuations originated from Refs.~\cite{evans:1993,gallavotti:1995a,gallavotti:1995b}, where a class of relations, now known as Fluctuation Relations (FR), was introduced, relating the probabilities of opposite energy dissipations of driven system. Close to equilibrium, FR reproduce the Green-Kubo and Onsager relations~\cite{gallavotti:1996,evans:2005}. Moreover, FR are among the few exact results valid almost arbitrarily far from equilibrium and have therefore attracted considerable interest~\cite{searles:2007,gomez-solano:2010,seifert:2012,ciliberto:2017}. Related relations have, in fact, been determined, for observables such as work heat and energy dissipation, in diverse frameworks~\cite{rondoni:2007,marini-bettolo-marconi:2008,gallavotti:2014-book,kurchan:2007,jakvsic:2011,evans:2002,seifert:2012,ciliberto:2017,evans:2016,dalcengio:2016,polettini:2017,wang:2011,wang:2013}, including dynamical systems and stochastic processes, classical and quantum systems, transient, steady-states and aging systems, for both global and local quantities, and for steady and time-dependent states. FR have also been experimentally verified in gravitational wave detectors~\cite{bonaldi:2009}.

The main ingredient to prove FR is some kind of time reversibility. For stochastic systems, this typically means detailed balance, while for deterministic dynamics this typically\footnote{Strict time-reversal invariance is not
required even in deterministic dynamics, as the FR are statistical relations~\cite{colangeli:2011,colangeli:2012}.} means the standard reversibility defined by the momentum inversion operator $\mathcal{M}_s : \mathfrak{M} \to \mathfrak{M}$:
\begin{equation}
\mathcal{M}_s (\vec{r},\vec{p}) = (\vec{r},-\vec{p}) ~, \quad \forall (\vec{r}, \vec{p}) \doteq \Gamma \in  \mathfrak{M}
\label{eq:oldsym}
\end{equation}
where $\Gamma$ is a point in the phase space $\mathfrak{M}$ of an $N$-particle system, with positions $\vec{r} = \{\vec{r}_i\}_{i=1}^{N}$ 
and momenta $\vec{p} = \{\vec{p}_i\}_{i=1}^{N}$. It is well known that the symmetry $\mathcal{M}_s$ is broken by an external magnetic field, $\vec{B}$. This has consolidated, also in the domain of FR, the idea that statistical properties of charged systems in external magnetic field necessitate special treatment. The usual approach extends the system to include the electric currents generating the magnetic field. Currents, and hence the magnetic field, are reversed under Eq.~\eqref{eq:oldsym} so the symmetry is restored, in the non-extended problem, by considering two systems subject to opposite external magnetic fields. Following this argument, Casimir~\cite{casimir:1945} modified the Onsager reciprocal relations to relate cross-transport coefficients of systems subject to $\vec{B}$ and $-\vec{B}$. Likewise, in his fundamental paper on linear response theory~\cite{kubo:1966}, Kubo established symmetry properties of time-correlation functions under the same conditions. In the context of FR, results for currents and non-equilibrium response were derived that also relate systems under opposite fields~\cite{gaspard:2013,barbier:2018a,wang:2014,saito:2008}. 
Unfortunately, this approach significantly limits the predictive power of the corresponding theories. For instance, identification of null values of transport coefficients in experiments concerning a single system in a given magnetic field based on symmetry is impossible. Similar considerations apply to systems rotating with constant angular velocity, where statistical relations involve two systems rotating with opposite angular velocities.

This point of view, adopted in classic textbooks~~\cite{landau:1980-book,degroot:1984-book}, is correct. However, observing that invariance of the Hamiltonian under Eq.~\eqref{eq:oldsym} is a sufficient but not necessary condition for the properties mentioned above, it was recently demonstrated~\cite{bonella:2014} that a more general approach is possible. There exist, in fact, alternative time-reversal operators~\cite{bonella:2017a,coretti:2018a} that, together with the change $t\rightarrow-t$, leave the evolving equations invariant without changing the sign of the magnetic field. Exploiting them, standard statistical relations can be immediately reinstated.  Refs. \cite{bonella:2017a,coretti:2018a} demonstrate this for time-correlation functions, illustrating the result with numerical simulations~\cite{bonella:2017a}. Consistently, no experimental evidence of the violation of the Onsager Reciprocal Relations is known~\cite{luo:2020}.

Here, we extend this single-system description, to transient and steady-state FR and to their corollaries, such as relations linking cumulants of currents to driving dissipative forces.

\section{Theory}
For convenience, we start by summarizing the derivation of transient and steady-state FR for general systems, stressing the role of time-reversal symmetry, which is detailed in the Supplementary Material (SM). Complete derivations of the FR can be found e.g.\ in Refs.~\cite{searles:2007,jepps:2010,rondoni:2016}. 

\subsection{General Theory of FR}
\label{sec:FR-general}
Consider a point $\Gamma \in \mathfrak{M}$, evolving under the dynamical equation $\dot{\Gamma} = G(\Gamma)$, where $G:\mathfrak{M}\to\mathfrak{M}$ is a vector field. Once the initial state $\Gamma_0$ is specified, this equation admits the formal solution $\Gamma_t = \mathcal{U}_t\Gamma_0$ where $\mathcal{U}_t: \mathfrak{M}\to\mathfrak{M}$ is the propagator for a time $t\in\mathbb{R}$. For any observable $\Psi: \mathfrak{M} \to \mathbb{R}$ and time interval $[t, t+\tau]$ with $\tau > 0$ we define
\begin{equation}
\label{eq:averages}
\Psi_{t,t+\tau}(\Gamma) \doteq \int_{t}^{t+\tau}\dd s\Psi(\mathcal{U}_s\Gamma)
\end{equation}
which is also an observable. The time average over a time $\tau$ of $\Psi$ is given by $\overline{\Psi}_{t,t+\tau}(\Gamma) \doteq \tau^{-1}\Psi_{t,t+\tau}(\Gamma)$.
For any interval $(a,b) \subset \mathbb{R}$ we denote by $\{\Psi\}_{(a,b)}$ the set of phase-space points such that $\Psi$ takes values in $(a,b)$:
\begin{equation*}
\mathfrak{M} \supset \{\Psi\}_{(a,b)} \doteq \{\Gamma \in \mathfrak{M} : \Psi(\Gamma) \in (a,b)\}
\end{equation*}
Let $\mathfrak{M}$ be endowed with a probability measure $\mu_0$ of density $f_0$, at time $t=0$, so that $\dd\mu_0(\Gamma) = f_0(\Gamma)\dd\Gamma$ is the probability of an infinitesimal volume element around $\Gamma$. The probability of finding the value of $\Psi$ in a given interval $(a,b)$ at time $t=0$ is given by
\begin{equation*}
\mu_0(\{\Psi\}_{(a,b)}) = \int_{\{\Psi\}_{(a,b)}}\!\!\!\!\!\!\!\!\!\!\!\!\!\!\dd\mu_0(\Gamma) = \int_{\{\Psi\}_{(a,b)}}\!\!\!\!\!\!\!\!\!\!\!\!\!\!f_0(\Gamma)\dd\Gamma
\end{equation*}
Assuming $f_0 \neq 0$ in $\mathfrak{M}$, the dissipation function $\Omegazero$ is 
\begin{equation}
\label{eq:Omegazero}
\Omega^{(0)}(\Gamma) \doteq -\nabla_\Gamma \ln f_0\cdot G(\Gamma) - \Lambda(\Gamma)
\end{equation}
where $\Lambda = \nabla\cdot\dot{\Gamma}$ is the phase-space expansion rate. 
An involution $\mathcal{M}:\mathfrak{M}\to\mathfrak{M}$ is a time-reversal symmetry if
\begin{equation}
\label{eq:TRI}
\mathcal{U}_{-t} \Gamma = \mathcal{M} \mathcal{U}_t \mathcal{M} \Gamma \quad
\forall t \in \mathbb{R} \ , \ \forall \Gamma \in \mathfrak{M}
\end{equation}
Assuming $f_0$ even under the action of $\mathcal{M}$, $f_0(\mathcal{M}\Gamma) = f_0(\Gamma)$, it is easy to show that the dissipation function is odd: $\Omegazero(\mathcal{M}\Gamma) = -\Omegazero(\Gamma)$. 

To derive the transient FR, consider the ratio of the initial probabilities to find the time average of $\Omegazero$ over $\tau$ in a neighborhood of size $\delta$ of $A$ and of $-A$ \cite{searles:2007,rondoni:2007}:
\begin{equation}
\label{eq:t-FR-init}
\begin{aligned}
\frac{\mu_0(\{\overline{\Omegazero}_{0,\tau}\}_{(-A)_{\delta}})}{\mu_0(\{\overline{\Omegazero}_{0,\tau}\}_{(A)_{\delta}})} = \frac{\int_{\{\overline{\Omegazero}_{0,\tau}\}_{(-A)_{\delta}}}f_0(\Gamma)\dd\Gamma}{\int_{\{\overline{\Omegazero}_{0,\tau}\}_{(A)_{\delta}}}f_0(\Gamma)\dd\Gamma}
\end{aligned}
\end{equation}
where we introduced the intervals $(\pm A)_{\delta} = (\pm A-\delta, \pm A+\delta) \subset \mathbb{R}$. Invoking the parity of $f_0$ under $\mathcal{M}$ and the relation between subsets of phase-space
\begin{equation}
\label{eq:SubSetRel}
\{\overline{\Omegazero}_{0,\tau}\}_{(-A)_{\delta}} = \mathcal{M}\mathcal{U}_\tau\{\overline{\Omegazero}_{0,\tau}\}_{(A)_{\delta}}
\end{equation}
Eq.~\eqref{eq:t-FR-init} can be written as
\begin{equation}
\label{eq:t-FR-fin}
\begin{aligned}
\frac{\mu_0(\{\overline{\Omega^{(0)}}_{0,\tau}\}_{(-A)_{\delta}})}{\mu_0(\{\overline{\Omega^{(0)}}_{0,\tau}\}_{(A)_{\delta}})} 
= \exp\Bigl[-\tau[A + \epsilon(\delta, A, \tau)]\Bigr]
\end{aligned}
\end{equation}
where $\epsilon$ is a correction term obeying $|\epsilon(\delta, A, \tau)| \leq \delta$. 
Eq.~\eqref{eq:t-FR-fin} is the transient FR, where ``transient'' means that it expresses a property of an initial state that is not stationary under the dynamics determined by the vector field $G$. In the SM, we show that Eq.~\eqref{eq:SubSetRel} is a direct consequence of time-reversal invariance of the dynamical system under $\mathcal{M}$. Thus, time-reversal invariance of the dynamics and of $f_0$ are the only requirements for the proof: the specific form of $\cal M$ is irrelevant, as long as Eq.~\eqref{eq:TRI} is satisfied.

Introducing the evolved probability measure $\mu_t$, defined by the conservation of probability $\mu_t(E) = \mu_0(\mathcal{U}_{-t}E)$, $E \subset \mathfrak{M}$, and taking the $t\to\infty$ limit followed by the $\tau\rightarrow\infty$ limit of Eq.\eqref{eq:t-FR-fin}, one may write~\cite{searles:2007,rondoni:2007}:
\begin{equation} 
\lim_{\tau \to \infty} \frac{1}{\tau} \ln
\frac{\mu_{\infty}(\{\overline{\Omega^{(0)}}_{0,\tau}\}_{(-A)_{\delta}})}
{\mu_{\infty}(\{\overline{\Omega^{(0)}}_{0,\tau}\}_{(A)_{\delta}})} = -[A +
\epsilon(A,\delta) - {\cal C}_0(A,\delta)] ~,
\label{exactFT2} 
\end{equation}
where $\mu_\infty(E)=\lim_{t\to\infty} \mu_t(E)$, $|\epsilon(A,\delta)|\leq \delta$, and
\begin{equation}
\label{eq:condautoc}
{\cal C}_0(A,\delta) \doteq \lim_{\tau\to\infty} \frac{1}{\tau} \lim_{t\to\infty} \left\langle e^{-\Omega^{(0)}_{0,t} - \Omega^{(0)}_{t+\tau,2t+\tau}} \right\rangle^{(0)}_{\{\overline{\Omega^{(0)}}_{t,t+\tau}\}_{(A)_{\delta}}}  
\end{equation}
with $\langle \cdot \rangle^{(0)}_{\{\overline{\Omega^{(0)}}_{t,t+\tau}\}_{(A)_{\delta}}}$ denoting an average with respect to $\mu_0$, under the condition $\overline{\Omega^{(0)}}_{t,t+\tau}(\Gamma) \in {(A)_\delta}$. Under the additional hypothesis that ${\cal C}_0(A,\delta)$ vanishes, Eq.~\eqref{exactFT2} represents the steady-state ($\mu_\infty$) FR. That correlations behave in such a way that ${\cal C}_0(A,\delta)$ vanishes is a non-trivial requirement. There are indeed systems that remain indefinitely trapped and do not reach a steady state.  

\subsection{Fluctuations Relations for $\vec{B}\neq0$}
\label{subsec:FR-magnetic}
Let us now consider a three-dimensional system of $N$ particles of charge $q_i$ and mass $m_i$, subject to uniform and static electric and magnetic fields, in a volume $\mathcal{V}$. The Hamiltonian is
\begin{equation}
\begin{aligned}
\label{eq:EM_Ham}
H(\Gamma) &= H_0(\Gamma) - \sum_{i=1}^Nq_i\vec{E}\cdot\vec{r}_i = \\
&= \sum_{i=1}^N\frac{\bigl(\vec{p}_i - q_i\vec{A}(\vec{r}_i)\bigr)^2}{2m_i} + \sum_{i,j<i}^NV(r_{ij}) - \sum_{i=1}^Nq_i\vec{E}\cdot\vec{r}_i
\end{aligned}
\end{equation}
where $\vec{A}(\vec{r})$ is the vector potential associated to the magnetic field $\vec{B} = \vec{\nabla} \times \vec{A}(\vec{r})$, $\vec{E}$ is the electric field and $V(r_{ij})$ is a pairwise additive interaction potential, depending only on the modulus of the distance between particles: $r_{ij} = |\vec{r}_i - \vec{r}_j|$. We orient the fields as $\vec{E} = (E_x, 0, 0)$ and $\vec{B} = (0, 0, B_z)$. A compatible vector potential, enforcing the Coulomb gauge $\vec{\nabla}_{\vec{r}} \cdot \vec{A}(\vec{r}) = 0$, is $\vec{A}(\vec{r}) = B_z/2(-y, x, 0)$. This setting, while not completely general, includes the majority of physically interesting cases and is usually adopted to discuss the time-reversal properties of systems in external magnetic fields~\cite{gaspard:2013,barbier:2018a,jayannavar:2007,poria:2016}.

We now consider deterministic thermostats coupled to this system. We present first time-reversal symmetries that make the proof of FR applicable, then we obtain explicit expressions for $\Omegazero$ and for the FR.

\subsubsection{The isokinetic non-equilibrium ensemble}\label{subsubsec:isokin}
The isokinetic thermostat is often used in connection with FR~\cite{searles:2007,searles:2013}. The isokinetic evolution associated to eq.~\eqref{eq:EM_Ham} is 
\begin{widetext}
\begin{equation}
\label{eq:PertEoM}
\begin{aligned}
\frac{\dd x_i}{\dd t} &= \frac{p^x_i}{m_i} + \omega_iy_i \\
\frac{\dd y_i}{\dd t} &= \frac{p^y_i}{m_i} - \omega_ix_i \\
\frac{\dd z_i}{\dd t} &= \frac{p^z_i}{m_i}
\end{aligned}
\qquad\qquad
\begin{aligned}
\frac{\dd p^x_i}{\dd t} &= F^x_i + \omega_i(p^y_i - m_i\omega_ix_i) + q_iE_x - \frac{\alpha_{\mathrm{IK}}}{2}(p^x_i + m_i\omega_iy_i)\\
\frac{\dd p^y_i}{\dd t} &= F^y_i - \omega_i(p^x_i + m_i\omega_iy_i) - \frac{\alpha_{\mathrm{IK}}}{2}(p^y_i - m_i\omega_ix_i)\\
\frac{\dd p^z_i}{\dd t} &= F^z_i - \frac{\alpha_{\mathrm{IK}}}{2}p^z_i
\end{aligned}
\end{equation}
\end{widetext}
where $F^{\lambda}_i$ and $\omega_i = \frac{B_z q_i}{2m_i}$ are the $\lambda$ Cartesian component of the interparticle force and the cyclotron frequency for particle $i$, respectively. Using Gauss' principle of least constraint (see SM), the thermostat parameter $\alpha_{\mathrm{IK}}$ is obtained as:
\begin{equation}
\label{eq:thermostat}
\begin{aligned}
\alpha_{\mathrm{IK}} &= \frac{\sum_{i=1}^N\vec{\Phi}_i\cdot\dot{\vec{r}}_i}{\frac{1}{2}\sum_{i=1}^Nm_i|\dot{\vec{r}}_i|^2} = \frac{\sum_{i=1}^N\vec{\Phi}_i\cdot(\vec{p}_i - q_i\vec{A}(\vec{r}_i))/m_i}{\sum_{i=1}^N|\vec{p}_i - q_i\vec{A}(\vec{r}_i)|^2/2m_i}
\end{aligned}
\end{equation}
where $\vec{\Phi}_i = -\vec{\nabla}_{\vec{r}_i} H$ are the active forces. Similar to previous studies~\cite{searles:2013}, we take $f_0$ as the equilibrium distribution
\begin{equation}
\label{eq:PSDensity}
f_0(\Gamma) = \frac{\exp\bigl[-\beta H_0(\Gamma)\bigr]\delta(K(\Gamma) - K^*)}{\int_{\mathfrak{M}}\dd\Gamma\exp\bigl[-\beta H_0(\Gamma)\bigr]\delta(K(\Gamma) - K^*)}
\end{equation}
(see SM). In the equation above, $H_0(\Gamma)$ is defined in Eq.~\eqref{eq:EM_Ham}, $K(\Gamma) = \sum_{i=1}^Nm_i|\dot{\vec{r}}_i|^2/2$ is the microscopic estimator of the kinetic energy ($K^*$ is the value fixed by the initial state) and $\beta = 1/k_{_B} T$. 
Direct inspection of Eq.~\eqref{eq:PertEoM} and Eq.~\eqref{eq:thermostat} shows that the dynamical system is invariant under the time-reversal transformations:
\begin{subequations}
\label{eq:St_TRS}
\begin{align}
\mathcal{M}^{(4)}\Gamma &= (x,-y,z,-p^x,p^y,-p^z) \\
\mathcal{M}^{(6)}\Gamma &= (x,-y,-z,-p^x,p^y,p^z)
\end{align}
\end{subequations}
(The superscripts reflect the nomenclature in Ref.~\cite{coretti:2018a} where both operators were  introduced.)
Inspection of Eq.~\eqref{eq:PSDensity}, shows that the initial probability density is even. The hypotheses introduced in Section~\ref{sec:FR-general} to derive of Eqs.~\eqref{eq:t-FR-fin} and~\eqref{exactFT2} are then satisfied and we can establish the explicit expression of the FR for this system. Note that the the validity of these time-reversal symmetries $\mathcal{M}^{(4)}$ and $\mathcal{M}^{(6)}$) depends on the orientation of the magnetic and electric fields. In Ref.~\cite{coretti:2018a}, however, it was shown that at least one time symmetry remains for arbitrary orientations of the fields, as long as the interparticle potentials is isotropic.

\subsubsection{The dissipation function and the fluctuation relations}
The explicit dissipation function is obtained by inserting the specific form of $f_0$, Eq.~\eqref{eq:PSDensity}, and of the equations of motion, Eq.~\eqref{eq:PertEoM}, in Eq.~\eqref{eq:Omegazero}. As shown in the SM, one obtains:
\begin{equation*}
\begin{aligned}
\Omega^{(0)}(\Gamma) = \beta \sum_{i=1}^N q_i\vec{E} \cdot \dot{\vec{r}}_i = \beta \mathcal{V} \vec{J}(\Gamma) \cdot \vec{E}
\end{aligned}
\end{equation*}
where the last equality defines the microscopic estimator for the electric current $\vec{J} = \mathcal{V}^{-1}\sum_{i=1}^Nq_i\dot{\vec{r}}_i$. 
The time-averaged dissipation function is  obtained, from Eq.~\eqref{eq:averages}, as $\overline{\Omega^{(0)}}_{0,\tau} = \beta {\cal V} \overline{\vec{J}}_{0,\tau}(\Gamma) \cdot \vec{E}$.
The dissipation function is proportional to the dissipative flux, hence to the dissipated energy. Moreover, as expected, $\Omega^{(0)}$ is odd under $\mathcal{M}^{(4)}$ and $\mathcal{M}^{(6)}$. 
The transient FR is obtained substituting in Eq.~\eqref{eq:t-FR-fin}:
\begin{equation}
\label{eq:t-FR-final}
\frac{\mu_0(\{ \beta {\cal V} \overline{\vec{J}}_{0,\tau} \cdot \vec{E} \}_{(-A)_{\delta}})}{\mu_0(\{ \beta {\cal V} \overline{\vec{J}}_{0,\tau} \cdot \vec{E} \}_{(A)_{\delta}})} = \exp\Bigl[-\tau[A + \epsilon(\delta, A, \tau)]\Bigr]
\end{equation}
For the steady-state FR to hold, ${\cal C}_0(A,\delta)$ of Eq.~\eqref{eq:condautoc} must vanish. Numerical findings show that the steady-state FR typically holds in 
chaotic particle systems, characterized by fast decay of correlations  \cite{rondoni:2007,marini-bettolo-marconi:2008}. In Ref.~\cite{searles:2013}, the test is explicitly performed for color diffusion, but it has never been done for systems in a magnetic field. While interparticle interactions promote disorder, hence decay of correlations, the Lorentz force tends to induce ordered circular motions that may hinder the decay of ${\cal C}_0(A,\delta)$. However, such an ordering effect may not be critical, as illustrated by the following example of non-interacting charged particles in constant external magnetic and electric fields oriented as in Eq.~\eqref{eq:EM_Ham}. In the absence of thermostat, this model is analytically solvable and yields
\begin{widetext}
\begin{equation*}
{\cal C}_0(A,\delta) = \lim_{\tau\to\infty} \frac{1}{\tau} \lim_{t\to\infty} \left\langle \exp\biggl[-\beta\sum_i^N\frac{E_xq_i{v^\perp_i}}{\omega_i}\Bigl\{\Upsilon_i(t)\right. + \left.2\sin\Bigl(\frac{\omega_i t}{2}\Bigr)\Bigl[\Theta_i(t)\cos(\omega_i\tau) + \Xi_i(t)\sin(\omega_i\tau)\Bigr]\Bigr\}\biggr]\right\rangle^{(0)}_{\{\overline{\Omegazero}_{t,t+\tau}\}_{(A)_{\delta}}}
\end{equation*}
\end{widetext}
where
\begin{equation*}
\begin{aligned}
\Upsilon_i(t) &= \cos(\phi_i)\bigl[1 - \cos(\omega_i t)\bigl] + \sin(\phi_i)\sin(\omega_i t) \\ 
\Theta_i(t) &= \cos(\phi_i)\sin\Bigl(\frac{3\omega_i t}{2}\Bigr) - \sin(\phi_i)\cos\Bigl(\frac{3\omega_i t}{2}\Bigr) \\
\Xi_i(t) &= \cos(\phi_i)\cos\Bigl(\frac{3\omega_i t}{2}\Bigr) - \sin(\phi_i)\sin\Bigl(\frac{3\omega_i t}{2}\Bigr) 
\end{aligned}
\end{equation*}
and $v_i^\perp$ and $\phi_i$ are constants fixed by the initial conditions and by the relative intensities of the fields (see SM for details).
Notably, the expression in angular brackets is bounded for all values of $t$ and $\tau$ implying that ${\cal C}_0(A,\delta)$ for this model is indeed zero. 
The thermostatted solution can be obtained numerically. As shown in SM, for appropriate relative intensities of the fields, the motion remains bounded in the direction parallel to the electric field, cancelling the correlation term also for a non-interacting isokinetic model. This analysis holds in general for the components of the electric field orthogonal to the magnetic field. If the fields have a parallel component, the magnetic field, which only influences the orthogonal motion, does not affect directly dissipation in the parallel direction.
Since interactions should further reduce correlation times, this argument suggests that the steady-state condition can be verified. Future studies will investigate more general situations.

Assuming convergence of Eq.~\eqref{eq:condautoc} and apart for an error $O(\tau^0)$ in the exponential, the steady-state FR can be written as
\begin{equation}
\label{eq:ss-FR-final}
\frac{\mu_\infty(\{ \beta {\cal V} \overline{\vec{J}}_{0,\tau} \cdot \vec{E} \}_{(-A)_{\delta}})}{\mu_\infty(\{ \beta {\cal V} \overline{\vec{J}}_{0,\tau} \cdot \vec{E} \}_{(A)_{\delta}})} = \exp\Bigl[-\tau[A + \epsilon(\delta, A, \tau)]\Bigr]
\end{equation}
where $|\epsilon(\delta, A, \tau)| \leq \delta$.

It is worth stressing that, although Eqs.~\eqref{eq:t-FR-final} and~\eqref{eq:ss-FR-final} are misleadingly similar, they refer to very different situations. Transient FR are associated to the statistics of the ensemble describing the initial (typically equilibrium) state. They describe a statistical property of many experiments of (short or long) duration $\tau$. Differently, steady-state FR refer to the steady state statistics of the currents of a single object or realization of the system. They require a kind of decorrelation between initial and final macrostates, which is why $t$ has to become large before $\tau$ does. This is not the mixing condition of ergodic theory, which corresponds to decay of correlations of microscopic events {\em within} a steady state.~\cite{rondoni:2007,marini-bettolo-marconi:2008,colangeli:2014}. 
If correlations do not decay, some kind of FR may still hold, but they (and derived relations) would take a different form, see {\em e.g.}\ Refs.~\cite{vanzon:2003,rondoni:2003,bonetto:2006,jepps:2010}.

\subsubsection{The generalized Nos\'e-Hoover thermostat}\label{subsubsec:NoseHoover}
At equilibrium, the Nos\'e-Hoover thermostat samples the canonical ensemble, and so does its generalization to systems in a magnetic field $\vec{B}$~\cite{mouhat:2013}. As for the case $\vec{B} = 0$, this generalization is based on the extension of the phase space through conjugate variables $s$ and $\xi$, mimicking the effect of a thermal bath. This thermostat can be easily modified to include a (static) external electric field (see also discussion in~\cite{mouhat:2013}), which allows us to extend the applicability of FR. The resulting generalized Nos\'e-Hoover dynamical system is:
\begin{widetext}
\begin{equation}
\label{eq:NH-EoM}
\begin{aligned}
\frac{\dd x_i}{\dd t} &= \frac{p^x_i}{m_i} + \omega_iy_i \\
\frac{\dd y_i}{\dd t} &= \frac{p^y_i}{m_i} - \omega_ix_i \\
\frac{\dd z_i}{\dd t} &= \frac{p^z_i}{m_i} \\
\frac{\dd \ln s}{\dd t} & = \xi
\end{aligned}
\qquad\qquad
\begin{aligned}
\frac{\dd p^x_i}{\dd t} &= F^x_i + \omega_i(p^y_i - m_i\omega_ix_i) + q_iE_x - \xi(p^x_i + m_i\omega_iy_i)\\
\frac{\dd p^y_i}{\dd t} &= F^y_i - \omega_i(p^x_i + m_i\omega_iy_i) - \xi(p^y_i - m_i\omega_ix_i)\\
\frac{\dd p^z_i}{\dd t} &= F^z_i - \xi p^z_i \\
\frac{\dd \xi}{\dd t} &= \frac{1}{\tau^2_{\mathrm{NH}}}\biggl[\frac{K(\Gamma) - K^*}{K^*}\biggr] = \frac{\delta K(\Gamma)}{\tau^2_{\mathrm{NH}}}
\end{aligned}
\end{equation}
\end{widetext}
where $\tau_{\mathrm{NH}}$ is the characteristic time of the thermostat. It is important to note that the kinetic energy of this system now fluctuates around the target value $K^*$, related to the temperature of the system via $\beta = 3N/(2K^*)$. As proved in~\cite{mouhat:2013}, the dynamical system \eqref{eq:NH-EoM} with $E_x=0$ conserves the quantity $H_{\mathrm{NH}}(\Gamma, \xi, s) = H(\Gamma) + K^*\bigl[\tau^2_{\mathrm{NH}}\xi^2 + 2\ln s\bigr]$ and samples the equilibrium distribution
\begin{equation}\label{eq:ExtCanonicalDens}
f_0(X) = \mathcal{Z}^{-1}\exp[-\beta H_0(\Gamma)]\exp[-\beta K^*\tau^2_{\mathrm{NH}}\xi^2]
\end{equation}
where $\mathcal{Z}$ is the partition function 
and $X$ denotes the extended phase-space $X = (\Gamma, \xi)$. As in standard Nos\'e-Hoover dynamics, the marginal probability obtained integrating Eq.~\eqref{eq:ExtCanonicalDens} with respect to $\xi$ is the canonical density for the physical variables. 

Direct inspection shows that \eqref{eq:NH-EoM} is invariant under  
\begin{subequations}
\label{eq:NH_TRS}
\begin{align}
\mathcal{M}_{\mathrm{ext}}^{(4)}(\Gamma, s, \xi) &= (x,-y,z,-p^x,p^y,-p^z,s,-\xi) \\
\mathcal{M}_{\mathrm{ext}}^{(6)}(\Gamma, s, \xi) &= (x,-y,-z,-p^x,p^y,p^z,s,-\xi)
\end{align}
\end{subequations}
together with time inversion. The equilibrium density Eq.~\eqref{eq:ExtCanonicalDens} is even under these transformations. The conditions for the transient FR are then verified and we can calculate the dissipation function, Eq.~\eqref{eq:Omegazero}. In the same fashion as the isokinetic case (see SM), it is possible to show that
$\nabla_X \ln f_0\cdot\dot{X} = \beta 2K^*\xi - \beta \sum^N_{i=1}q_i\dot{\vec{r}}_i\cdot\vec{E} - \beta \xi\delta K(\Gamma)$
while the compressibility of the (extended) phase space is given by $\Lambda = -\beta 2K^*\xi$. Substituting in  Eq.~\eqref{eq:Omegazero} we obtain
\begin{equation}
\Omegazero(X) = \beta V \vec{J}(\Gamma)\cdot\vec{E} + \beta \xi\delta K(\Gamma)
\end{equation} 
for the \emph{instantaneous} dissipation function of the system 
\eqref{eq:NH-EoM}, odd under the valid time-reversal symmetries.
There are now two sources of dissipation: the electric field and the temperature gradient between system and reservoir.
In the expression for the average dissipation function $\overline{\Omega^{(0)}}_{0,\tau}$, the contribution due to the temperature gradient is negligible compared to the other, for $\tau \gg \tau_{\mathrm{NH}}$. In this limit the FR take the same form as for the isokinetic case.

\section{Concluding remarks}
We have shown that transient and steady-state FR can be derived in
the presence of a static and uniform magnetic field, without inversion of $\vec{B}$. This is possible because the dynamical system admits time-reversal symmetries that, at variance with the standard momentum reversal, are not violated by the field. Steady-state FR require, as always, the decay of appropriate correlations. For $\vec{B}=0$, this condition may be violated under strong drivings inducing ordered phases, in which back currents are suppressed~\cite{lloyd:1995,rondoni:2003,bonetto:2006}. The effect of magnetic fields on these correlations needs further investigation, but in the case discussed above they do not alter the validity of the FR.

Use of a single magnetic field immediately improves the predictive power of the theory. For instance, consider a vector of $n$ affinities $\vec{\mathrm{A}}$, the corresponding $n$ amounts of energy and matter exchanged between the reservoirs and a reference reservoir in a time interval $[0, t]$, $\Delta \vec{x}$, and an $n$-dimensional vector of parameters $\vec{\lambda}$.
Let the cumulant generating function of $\vec{\mathrm{A}}$ be defined by:
\begin{equation*}
G_t(\vec{\lambda},\vec{\mathrm{A}};\vec{B}) = \int p_t(\Delta \vec{x},\vec{\mathrm{A}};\vec{B}) \exp\left(-\vec{\lambda}\cdot\Delta\vec{x}\right)
\dd\Delta\vec{x}
\end{equation*}
where $p_t$ is the probability density of $\Delta\vec{x}$ derived from the grand-canonical ensemble, at fixed affinities and constant $\vec{B}$. Then, following the procedure for asymptotic (not necessarily steady-state) FR~\cite{searles:2007}, Ref.~\cite{barbier:2018a} defines the asymptotic generating function as $Q(\vec{\lambda},\vec{\mathrm{A}};\vec{B}) = -\lim_{t \to \infty} (1/t) \ln G_t(\vec{\lambda},\vec{\mathrm{A}};\vec{B})$.
The corresponding cumulants, {\em i.e.}\ the derivatives of $Q$ with respect to the components of $\vec{\lambda}$
evaluated at $\vec{\lambda}=\vec{0}$, are then expanded as power series of $\vec{\mathrm{A}}$, around $\vec{\mathrm{A}}=\vec{0}$:
\begin{equation*}
Q(\vec{\lambda},\vec{\mathrm{A}};\vec{B}) = \sum_{m,n=0}^\infty 
\frac{Q_{\alpha_1\dots\beta_n}(\vec{B})}{m!n!} \lambda_{\alpha_1} \cdots \lambda_{\alpha_m}
\mathrm{A}_{\beta_1} \cdots \mathrm{A}_{\beta_n}
\end{equation*}
with
$\lambda_i$ the $i$-th element of $\vec{\lambda}$, $\mathrm{A}_j$ the $j$-th affinity and:
\begin{equation*}
Q_{\alpha_1\dots \beta_n}({\vec{B}}) = \left.\frac{\partial^{m+n} Q}{\partial \lambda_{\alpha_1} \cdots \partial \lambda_{\alpha_m} \partial \mathrm{A}_{\beta_1} \cdots \partial \mathrm{A}_{\beta_n}} \right|_{\vec{\lambda}=\vec{0};\vec{\mathrm{A}}=\vec{0}}
\end{equation*}
In terms of $Q$ and of the reversibility based on the inversion of $\vec{B}$, the asymptotic FR is then written as $Q(\vec{\lambda},\vec{\mathrm{A}};\vec{B}) = Q(\vec{\mathrm{A}} - \vec{\lambda},\vec{\mathrm{A}};-\vec{B})$ which imposes certain constraints on $Q_{\alpha_1...\beta_n}$. For instance, Eq.~(43) of Ref.~\cite{barbier:2018a} states that $Q_{\alpha_1 \cdots \alpha_m}(\vec{0};\vec{B}) = 
(-1)^m Q_{\alpha_1 \cdots \alpha_m}(\vec{0};-\vec{B})$.
Using instead ${\cal M}^{(4)}$ or ${\cal M}^{(6)}$ of Eqs.~\eqref{eq:St_TRS}, one also obtains $Q_{\alpha_1 \cdots \alpha_m}(\vec{0};\vec{B}) = 
(-1)^m Q_{\alpha_1 \cdots \alpha_m}(\vec{0};\vec{B}) 
$ which entails the stronger result $Q_{\alpha_1 \cdots \alpha_m}(\vec{0};\vec{B}) = 0$ for odd $m$ and any $\vec{B}$.

The work presented in this paper thus enables a reformulation of general results, based in particular on FR {\em e.g.}\ Refs.~\cite{searles:2007,barbier:2018a}, lifting the prescription of opposite magnetic fields (or angular velocities) and restores the full predictive power of a number of statistical results for systems long considered as exceptions.

\begin{acknowledgments}
AC and LR have been partially supported by Ministero dell'Istruzione, dell'Universit\`{a} e della Ricerca 
(MIUR) grant ``Dipartimenti di Eccellenza 2018-2022''. 
\end{acknowledgments}

%
%
%

\pagebreak
\onecolumngrid
\large
\patchcmd{\large}{15}{15}{}{}
\begin{center}
  \textbf{\LARGE Supplementary material: Fluctuation relations for systems in constant magnetic field}\\[.2cm]
  Alessandro Coretti,$^{1,2}$ Lamberto Rondoni,$^{1,3}$ and Sara Bonella$^{2,*}$\\[.1cm]
  {\itshape ${}^1$Department of Mathematical Sciences, Politecnico di Torino, Corso Duca degli Abruzzi 24, I-10129 Torino, Italy\\
  ${}^2$Centre Europ\'een de Calcul Atomique et Mol\'eculaire (CECAM), \'Ecole Polytechnique F\'ed\'erale de Lausanne, Batochime, Avenue Forel 2, 1015 Lausanne, Switzerland\\
  ${}^3$Istituto Nazionale di Fisica Nucleare, Sezione di Torino, Via P. Giura 1, I-10125 Torino, Italy\\
  }
  ${}^*$Electronic address: sara.bonella@epfl.ch\\
(Dated: \today)\\[2cm]
\end{center}

\setcounter{equation}{0}
\setcounter{figure}{0}
\setcounter{table}{0}
\setcounter{page}{1}
\setcounter{section}{0}
\renewcommand{\theequation}{S\arabic{equation}}
\renewcommand{\thefigure}{S\arabic{figure}}
\renewcommand{\bibnumfmt}[1]{[S#1]}
\renewcommand{\citenumfont}[1]{S#1}
\renewcommand{\thesection}{S\Roman{section}}
\renewcommand{\thepage}{S\arabic{page}}

\titleformat*{\section}{\Large\bfseries}

\section{Proof that $\{\overline{\Omegazero}_{0,\tau}\}_{(-A)_{\delta}} = \mathcal{M}\mathcal{U}_\tau\{\overline{\Omegazero}_{0,\tau}\}_{(A)_{\delta}}$} \label{app:operator_on_set}
For any  observable that is odd under the chosen time reversal symmetry, $\Psi(\mathcal{M}\Gamma) = - \Psi(\Gamma)$, we prove the identity
\begin{equation*}
\{\overline{\Psi}_{0,\tau}\}_{(-A)_{\delta}} = \mathcal{M}\mathcal{U}_\tau\{\overline{\Psi}_{0,\tau}\}_{(A)_{\delta}}
\end{equation*}
establishing that phase-space points belonging to the subset in the LHS of the equation are those and only those that also belong to the set on the RHS. Setting $\Psi(\Gamma)=\Omegazero(\Gamma)$ demonstrates the result used in in Section IIA of the main text.

Let us begin by establishing the action of $\mathcal{M}\mathcal{U}_\tau$ on a subset of phase space. We have (see definition of $\{\Psi\}_{(a,b)}$ in the main text)
\begin{equation}
\label{eq:equality_set}
\begin{aligned}
\mathcal{M}\mathcal{U}_\tau\{\Psi\}_{(a,b)} &= \mathcal{M}\mathcal{U}_\tau\{\Gamma \in \mathfrak{M} : \Psi(\Gamma) \in (a,b)\} = \\
&= \{(\Gamma' = \mathcal{M}\mathcal{U}_\tau\Gamma) \in \mathfrak{M} : \Psi(\Gamma) \in (a,b)\} = \\
&= \{\Gamma' \in \mathfrak{M} : \Psi(\Gamma = (\mathcal{M}\mathcal{U}_\tau)^{-1}\Gamma') \in (a,b)\} = \\
&= \{\Gamma' \in \mathfrak{M} : \Psi(\mathcal{U}_{-\tau}\mathcal{M}\Gamma') \in (a,b)\}
\end{aligned}
\end{equation}
In the last equality, we used the properties $\mathcal{U}_\tau^{-1}=\mathcal{U}_{-\tau}$ and $\mathcal{M}^{-1}=\mathcal{M}$.

We will now show that, for any $\Gamma \in \{\overline{\Psi}_{0,\tau}\}_{(-A)_{\delta}}$, then $\mathcal{M}\mathcal{U}_\tau\Gamma \in \{\overline{\Psi}_{0,\tau}\}_{(A)_{\delta}}$. Let us consider the expression for the $\overline{\Psi}_{0,\tau}(\mathcal{U}_{-\tau}\mathcal{M}\Gamma)$ that, based on Eq.~(2) of the main text and on Eq.~\eqref{eq:equality_set}, is the value of the time-averaged observable on a point of the transformed phase-space subset. We have
\begin{equation*}
\overline{\Psi}_{0,\tau}(\mathcal{U}_{-\tau}\mathcal{M}\Gamma) = \frac{1}{\tau}\int_0^\tau \dd s\Psi(\mathcal{U}_s\mathcal{U}_{-\tau}\mathcal{M}\Gamma) = \frac{1}{\tau}\int_0^\tau \dd s\Psi(\mathcal{U}_{s-\tau}\mathcal{M}\Gamma)
\end{equation*}
where, in the last equality, the time-composition property of the propagator was employed. Performing the change of variable $t=s-\tau$, the integral becomes
\begin{equation}
\label{eq:int_dt}
\overline{\Psi}_{0,\tau}(\mathcal{U}_{-\tau}\mathcal{M}\Gamma) = \frac{1}{\tau}\int_{-\tau}^0 \dd t\Psi(\mathcal{U}_{t}\mathcal{M}\Gamma)
\end{equation}
Using the definition of time-reversal symmetry, Eq.~(6) in the main text, we have that $\mathcal{U}_t\mathcal{M}=\mathcal{M}\mathcal{U}_{-t}$ and Eq.~\eqref{eq:int_dt} can be written as
\begin{equation*}
\overline{\Psi}_{0,\tau}(\mathcal{U}_{-\tau}\mathcal{M}\Gamma) = \frac{1}{\tau}\int_{-\tau}^0 \dd t\Psi(\mathcal{M}\mathcal{U}_{-t}\Gamma)
\end{equation*}
Another change of the integration variable $u = -t$ and the exchange of the integration extrema now yield
\begin{equation}
\label{eq:int_du}
\overline{\Psi}_{0,\tau}(\mathcal{U}_{-\tau}\mathcal{M}\Gamma) = \frac{1}{\tau}\int_{0}^\tau  \dd u\Psi(\mathcal{M}\mathcal{U}_{u}\Gamma)= -\frac{1}{\tau}\int_{0}^\tau \dd u\Psi(\mathcal{U}_{u}\Gamma)= - 
\overline{\Psi}_{0,\tau}(\Gamma)
\end{equation}
In going from the second to the third equality, the odd parity of the observable was used, while the last equality recognizes the definition of Eq.~(2) in the main text. From Eq.~\eqref{eq:int_du}, it immediately follows that if $\overline{\Psi}_{0,\tau}(\mathcal{U}_{-\tau}\mathcal{M}\Gamma) \in [-A-\delta,-A+\delta]$ then $\overline{\Psi}_{0,\tau}(\Gamma) \in [A-\delta,A+\delta]$ and viceversa for any phase-space point $\Gamma$, which completes the proof.

\section{Derivation of the isokinetic thermostat parameter in constant magnetic field}
The standard derivation of the isokinetic thermostat parameter uses Gauss' principle of least constraint to obtain the equations of motion of the system minimizing the curvature 
\begin{equation}
\label{eq:GaussCurvature}
\mathcal{C} = \sum_{i=1}^Nm_i\biggl[\ddot{\vec{r}}_i - \frac{\vec{\Phi}_i}{m_i}\biggr]^2
\end{equation}
subject to the selected constraint(s). In eq.~\eqref{eq:GaussCurvature} the $\ddot{\vec{r}}_i$ represent the constrained accelerations, and $\vec{\Phi}_i$ the right hand sides of the unconstrained (non-equilibrium) dynamics expressed in Newtonian form
\begin{equation*}
\begin{split}
m_i\ddot{x}_i &= F^x_i + 2m_i\omega_i\dot{y}_i + q_iE_x\\
m_i\ddot{y}_i &= F^y_i - 2m_i\omega_i\dot{x}_i\\
m_i\ddot{z}_i &= F^z_i\\
\end{split}
\end{equation*}
The isokinetic constraint is given by
\begin{equation}
\label{eq:constraint}
g(\vec{r}, \dot{\vec{r}}) = K - K^* = \frac{1}{2}\sum_{i=1}^Nm_i|\dot{\vec{r}}_i|^2 - K^* = 0
\end{equation}
where $K$ is the standard microscopic estimator of the kinetic energy defined in Eq.~(13) in the main text and $K^*$ indicates the specific value of the kinetic energy set by the initial state.

To obtain an explicit expression for the constraint as a function of the variables $\ddot{\vec{r}}$ we take the derivative with respect to time of eq.~\eqref{eq:constraint}
\begin{equation}
\label{eq:dotconstraint}
\dot{g}(\vec{r}, \dot{\vec{r}}) = \sum_{i=1}^N \vec{\nabla}_{\vec{r}_i}g(\vec{r}, \dot{\vec{r}})\cdot\ddot{\vec{r}}_i = \sum_{i=1}^Nm_i\dot{\vec{r}}_i\cdot\ddot{\vec{r}}_i = 0
\end{equation}
and then we minimize, with respect to $\ddot{\vec{r}}_j$, the curvature subject to the constraint, leading to:
\begin{equation*}
\frac{d}{d\ddot{\vec{r}}_j}\Biggl\{\sum_{i=1}^Nm_i\biggl[\ddot{\vec{r}}_i - \frac{\vec{\Phi}_i}{m_i}\biggr]^2 - \alpha\sum_{i=1}^Nm_i\dot{\vec{r}}_i\cdot\ddot{\vec{r}}_i \Biggr\} = 0 \qquad j=1,\dots,N
\end{equation*}
where $\alpha$ is a Lagrange multiplier to be determined in order to satisfy the constraint. This yields the equations of motion for the constrained system
\begin{equation}
\label{eq:ConstrNewtEoM}
\begin{split}
m_i\ddot{\vec{r}}_i &= \vec{\Phi}_i - \alpha \frac{m_i}{2}\dot{\vec{r}}_i\\
\end{split}
\end{equation}
The Lagrange multiplier $\alpha$ is determined via eq.~\eqref{eq:dotconstraint} multiplying eq.~\eqref{eq:ConstrNewtEoM} by $\dot{\vec{r}}_i$ and summing over $i=1,\dots,N$
\begin{equation*}
\begin{split}
\sum_{i=1}^Nm_i\dot{\vec{r}}_i\ddot{\vec{r}}_i &= \dot{g}(\vec{r}, \dot{\vec{r}}) = 0 = \sum_{i=1}^N\biggl[\vec{\Phi}_i\cdot\dot{\vec{r}}_i - \alpha \frac{m_i}{2}\dot{\vec{r}}_i\cdot\dot{\vec{r}}_i\biggr]\\
\end{split}
\end{equation*}
to obtain
\begin{equation*}
\alpha_{\mathrm{IK}} = \frac{\sum_{i=1}^N\vec{\Phi}_i\cdot\dot{\vec{r}}_i}{\frac{1}{2}\sum_{i=1}^Nm_i|\dot{\vec{r}}_i|^2} = \frac{\sum_{i=1}^N\vec{\Phi}_i\cdot\dot{\vec{r}}_i}{K}\\
\end{equation*}
Substituting in Eq.~\eqref{eq:ConstrNewtEoM} and formulating the evolution in Hamiltonian form we obtain the dynamical system introduced in Section IIB1.

\section{Equilibrium distribution for the non-dissipative isokinetic ensemble} \label{app:eq_distribution}
Considering the equilibrium Hamiltonian $H_0(\Gamma)$ defined in Eq.~(10) in the main text, we show that 
\begin{equation*}
f_0(\Gamma) = \frac{\exp\bigl[-\beta H_0(\Gamma)\bigr]\delta(K(\Gamma) - K^*)}{\int_{\mathfrak{M}}\dd\Gamma\exp\bigl[-\beta H_0(\Gamma)\bigr]\delta(K(\Gamma) - K^*)}
\end{equation*}
is the equilibrium distribution (i.e. $\vec{E}=0$) for the isokinetic dynamics by verifying, via direct substitution, that it satisfies the generalized Liouville equation specialized to the equilibrium isokinetic dynamical system, Eq.~(11) in the main text with $\vec{E}=0$. Note that in presence of an external magnetic field, the total momentum is conserved on average, not instantaneously, explaining the lack of the delta function on momentum usually present in the isokinetic density (see, for example, Ref.~\cite{searles:2013}). At equilibrium, the generalized Liouville equation reads 
\begin{equation}
\label{eq:ss-liouville}
(\nabla_\Gamma f) \cdot \dot{\Gamma} + f(\nabla_\Gamma \cdot \dot{\Gamma}) = 0
\end{equation}
Let us consider first the compressibility of the system $\Lambda = \nabla_{\Gamma}\cdot\dot{\Gamma}$. To set the stage, we write the dynamical system in the form
\begin{equation*}
\begin{aligned}
\dot{\vec{r}}_i &= \vec{\grad}_{\vec{p}_i}H_0 \\
\dot{\vec{p}}_i &= -\vec{\grad}_{\vec{r}_i}H_0 - \frac{\alpha^0_\mathrm{IK}}{2}(\vec{p}_i - q_i\vec{A}(\vec{r}_i))
\end{aligned}
\end{equation*}
with
\begin{equation*}
\alpha^0_{\mathrm{IK}} = \frac{\sum_{i=1}^N\vec{\Phi}^0_i\cdot(\vec{p}_i - q_i\vec{A}(\vec{r}_i))/m_i}{\sum_{i=1}^N|\vec{p}_i - q_i\vec{A}(\vec{r}_i)|^2/2m_i} = \frac{1}{K}\sum_{i=1}^N\vec{\Phi}^0_i\cdot\frac{\vec{p}_i - q_i\vec{A}(\vec{r}_i)}{m_i}
\end{equation*}
where $\vec{\Phi}^0_i = -\vec{\grad}_{\vec{r}_i}H_0$.
The superscript $0$ indicates that we are considering the equilibrium situation with $\vec{E} = 0$. The compressibility can then be computed as
\begin{equation}
\label{eq:BeginComp}
\begin{aligned}
\nabla_{\Gamma}\cdot\dot{\Gamma} &= \sum_{i=1}^N\biggl\{\cancel{\vec{\grad}_{\vec{r}_i}\cdot\vec{\grad}_{\vec{p}_i}H_0} - \cancel{\vec{\grad}_{\vec{p}_i}\cdot\vec{\grad}_{\vec{r}_i}H_0} + \\
&- \vec{\grad}_{\vec{p}_i}\frac{\sum_{j=1}^N\vec{\Phi}^0_j\cdot(\vec{p}_j - q_j\vec{A}(\vec{r}_j))/m_j}{\sum_{j=1}^N|\vec{p}_j - q_j\vec{A}(\vec{r}_j)|^2/m_j}\cdot(\vec{p}_i - q_i\vec{A}(\vec{r}_i))\biggr\}
\end{aligned}
\end{equation}
Calculating the gradient in $\vec{p}_i$ yields
\begin{equation}
\label{eq:compr-one}
\begin{aligned}
\nabla_{\Gamma}\cdot\dot{\Gamma} &= - \frac{\sum_{j=1}^N\vec{\Phi}^0_j\cdot(\vec{p}_j - q_j\vec{A}(\vec{r}_j))/m_j}{\sum_{j=1}^N|\vec{p}_j - q_j\vec{A}(\vec{r}_j)|^2/m_j}\cancelto{3N}{\sum_{i=1}^N\vec{\grad}_{\vec{p}_i}\cdot(\vec{p}_i - q_i\vec{A}(\vec{r}_i))} + \\
&- \sum_{i=1}^N(\vec{p}_i - q_i\vec{A}(\vec{r}_i)) \cdot \vec{\grad}_{\vec{p}_i}\frac{\sum_{j=1}^N\vec{\Phi}^0_j\cdot(\vec{p}_j - q_j\vec{A}(\vec{r}_j))/m_j}{\sum_{j=1}^N|\vec{p}_j - q_j\vec{A}(\vec{r}_j)|^2/m_j}
\end{aligned}
\end{equation}
The gradient in the last term of the equation above can be written as
\begin{equation*}
\begin{aligned}
&\vec{\grad}_{\vec{p}_i}\frac{\sum_{j=1}^N\vec{\Phi}^0_j\cdot(\vec{p}_j - q_j\vec{A}(\vec{r}_j))/m_j}{\sum_{j=1}^N|\vec{p}_j - q_j\vec{A}(\vec{r}_j)|^2/m_j} = \frac{1}{4K^2}\biggl[2K\vec{\grad}_{\vec{p}_i}\sum_{j=1}^N\vec{\Phi}^0_j\cdot(\vec{p}_j - q_j\vec{A}(\vec{r}_j))/m_j + \\
&-\sum_{j=1}^N\vec{\Phi}^0_j\cdot(\vec{p}_j - q_j\vec{A}(\vec{r}_j))/m_j\vec{\grad}_{\vec{p}_i}\sum_{j=1}^N|\vec{p}_j - q_j\vec{A}(\vec{r}_j)|^2/m_j\biggr] = \\
& = \frac{1}{4K^2}\biggl[2K\vec{\Phi}^0_i/m_i -2(\vec{p}_i - q_i\vec{A}(\vec{r}_i))/m_i\sum_{j=1}^N\vec{\Phi}^0_j\cdot(\vec{p}_j - q_j\vec{A}(\vec{r}_j))/m_j\biggr]
\end{aligned}
\end{equation*}
Substituting in eq.~\eqref{eq:compr-one} yields
\begin{equation*}
\begin{aligned}
\nabla_{\Gamma}\cdot\dot{\Gamma} &= - 3N\frac{\sum_{j=1}^N\vec{\Phi}^0_j\cdot(\vec{p}_j - q_j\vec{A}(\vec{r}_j))/m_j}{\sum_{j=1}^N|\vec{p}_j - q_j\vec{A}(\vec{r}_j)|^2/m_j} - \frac{1}{4K^2}\sum_{i=1}^N(\vec{p}_i - q_i\vec{A}(\vec{r}_i)) \cdot \\
&\cdot \biggl[2K\vec{\Phi}^0_i/m_i -2(\vec{p}_i - q_i\vec{A}(\vec{r}_i))/m_i\sum_{j=1}^N\vec{\Phi}^0_j\cdot(\vec{p}_j - q_j\vec{A}(\vec{r}_j))/m_j\biggr] = \\
&= - 3N\frac{\sum_{j=1}^N\vec{\Phi}^0_j\cdot(\vec{p}_j - q_j\vec{A}(\vec{r}_j))/m_j}{\sum_{j=1}^N|\vec{p}_j - q_j\vec{A}(\vec{r}_j)|^2/m_j} + \\
&- \frac{1}{2K}\biggl[ \sum_{j=1}^N\vec{\Phi}^0_j\cdot(\vec{p}_j - q_j\vec{A}(\vec{r}_j))/m_j -2\sum_{j=1}^N\vec{\Phi}^0_j\cdot(\vec{p}_j - q_j\vec{A}(\vec{r}_j))/m_j\biggr] = \\
&=-\frac{\sum_{j=1}^N\vec{\Phi}^0_j\cdot(\vec{p}_j - q_j\vec{A}(\vec{r}_j))/m_j}{\sum_{j=1}^N|\vec{p}_j - q_j\vec{A}(\vec{r}_j)|^2/m_j}(3N - 1) = -\frac{\alpha^0_{\mathrm{IK}}}{2}(3N-1)
\end{aligned}
\end{equation*}
From the equipartition theorem for a system of $N$ particles in three dimensions with a frozen degree of freedom due to the isokinetic constraint we can write
\begin{equation*}
\frac{3N - 1}{2}k_BT = K \quad \Longrightarrow \quad \beta = \frac{3N - 1}{2K}
\end{equation*}
so that, finally,
\begin{equation}
\label{eq:CompLiouv}
\nabla_{\Gamma}\cdot\dot{\Gamma} = -\beta K \alpha^0_{\mathrm{IK}}
\end{equation}

Next we compute the gradient of the distribution
\begin{equation}
\label{eq:grad_gamma-one}
\begin{aligned}
\nabla_{\Gamma}f_0 &= \nabla_{\Gamma}\mathcal{Z}^{-1}\exp[-\beta H_0]\delta(K - K^*) \\
&=\mathcal{Z}^{-1}\Bigl[\delta(K - K^*)\nabla_{\Gamma}\exp[-\beta H_0] + \exp[-\beta H_0] \nabla_{\Gamma}\delta(K - K^*)\Bigr]
\end{aligned}
\end{equation} 
where $\mathcal{Z} = \int d\Gamma\exp[-\beta H_0]\delta(K - K^*)$. The first term in the last line above is easily computed as
\begin{equation}
\begin{aligned}
\mathcal{Z}^{-1}&\delta(K - K^*)\nabla_{\Gamma}\exp[-\beta H_0] = \\
& =\mathcal{Z}^{-1}\delta(K - K^*)\bigl(-\beta\exp[-\beta H^*]\nabla_\Gamma H_0\bigr) = -\beta f_0 \nabla_\Gamma H_0
\end{aligned}
\end{equation}
Scalar product of this term with the flux on the phase space (see first term in Eq.~\eqref{eq:ss-liouville}) gives
\begin{equation}
\label{eq:FirstTermGradEqDens}
\begin{aligned}
-\beta f_0\nabla_\Gamma H_0 \cdot \dot{\Gamma} &= \frac{\beta}{2} f \sum_{i=1}^N \vec{\nabla}_{\vec{p}_i}H_0 \cdot \alpha^0_{\mathrm{IK}}(\vec{p}_i - q_i\vec{A}(\vec{r}_i)) = \\
&= \frac{\beta}{2} f\sum_{i=1}^N \frac{\vec{p}_i - q_i\vec{A}(\vec{r}_i)}{m_i} \cdot \alpha^0_{\mathrm{IK}}(\vec{p}_i - q_i\vec{A}(\vec{r}_i)) = \beta f_0 K \alpha^0_{\mathrm{IK}}
\end{aligned}
\end{equation}
We shall now show that second term of the RHS of Eq.~\eqref{eq:grad_gamma-one} vanishes. We have
\begin{equation}
\label{eq:InitNull}
\begin{aligned}
\mathcal{Z}^{-1}\exp[-\beta H_0] \nabla_{\Gamma}\delta(K - K^*) = \mathcal{Z}^{-1}\exp[-\beta H_0] \delta'(K - K^*)\nabla_{\Gamma}K
\end{aligned}
\end{equation}
where we introduced the quantity
\begin{equation*}
\frac{d}{dx}\delta(x-x_0) = \delta'(x-x_0)
\end{equation*}
The chain rule for the Dirac's delta function can be easily obtained in the framework of the generalized functions~\cite{lighthill:1958-book}. The scalar product of the derivative above with the phase-space flux is given by
\begin{equation*}
\begin{aligned}
\nabla_{\Gamma}K\cdot\dot{\Gamma} = \sum_{i=1}^N\biggl[ \vec{\nabla}_{\vec{r}_i}K\vec{\nabla}_{\vec{p}_i}H_0 - \vec{\nabla}_{\vec{p}_i}K\vec{\nabla}_{\vec{r}_i}H_0 - \vec{\nabla}_{\vec{p}_i}K\cdot\frac{\alpha^0_{\mathrm{IK}}}{2}(\vec{p}_i - q_i\vec{A}(\vec{r}_i))\biggr]
\end{aligned}
\end{equation*}
The first term of the RHS is null since $\vec{\nabla}_{\vec{r}_i}K = 0$ as easily verified remembering that, in the adopted Coulomb gauge,  $\vec{\nabla}_{\vec{r}_i}\cdot\vec{A}(\vec{r}_i) = 0$. The other two terms in square parenthesis are equal and opposite. Indeed
\begin{equation*}
\begin{aligned}
-\sum_{i=1}^N \vec{\nabla}_{\vec{p}_i}K\vec{\nabla}_{\vec{r}_i}H_0 =  \sum_{i=1}^N\vec{\Phi}^0_i\cdot\frac{\vec{p}_i - q_i\vec{A}(\vec{r}_i)}{m_i}
\end{aligned}
\end{equation*}
and
\begin{equation}
\label{eq:FinNull}
\begin{aligned}
-\sum_{i=1}^N\vec{\nabla}_{\vec{p}_i}K\cdot\frac{\alpha^0_{\mathrm{IK}}}{2}(\vec{p}_i - q_i\vec{A}(\vec{r}_i)) &= -\alpha^0_{\mathrm{IK}}\sum_{i=1}^N\frac{|\vec{p}_i - q_i\vec{A}(\vec{r}_i)|^2}{2m_i} = \\
&= -\sum_{i=1}^N\vec{\Phi}^0_i\cdot\frac{\vec{p}_i - q_i\vec{A}(\vec{r}_i)}{m_i}
\end{aligned}
\end{equation}
Summarizing, the calculations above show that the non-zero contributions to the generalized Liouville equation arise from the compressibility, Eq.~\eqref{eq:CompLiouv}, and from Eq.~\eqref{eq:FirstTermGradEqDens}. Substituting in Eq.~\eqref{eq:ss-liouville} then completes the proof since 
\begin{equation*}
(\nabla_\Gamma f_0) \cdot \dot{\Gamma} + f_0(\nabla_\Gamma \cdot \dot{\Gamma}) = \beta f_0 K \alpha^0_{\mathrm{IK}} - \beta f_0 K \alpha^0_{\mathrm{IK}} = 0
\end{equation*}

\section{The dissipation function for the dissipative isokinetic ensemble}
The dissipation function 
\begin{equation*}
\Omega^{(0)}(\Gamma) = -\frac{1}{f_0}\nabla_\Gamma f_0\cdot\dot{\Gamma} - \nabla_\Gamma\cdot\dot{\Gamma}
\end{equation*}
is computed by direct substitution in the definition above of the equilibrium density and of the --- non-equilibrium --- phase-space flux, Eq.~(14) in the main text with $\vec{E}\neq 0$.
The relevant dynamical system is thus
\begin{equation}
\label{eq:NonEqHamEq}
\begin{aligned}
\dot{\vec{r}}_i &= \vec{\grad}_{\vec{p}_i}H \\
\dot{\vec{p}}_i &= -\vec{\grad}_{\vec{r}_i}H - \frac{\alpha_\mathrm{IK}}{2}(\vec{p}_i - q_i\vec{A}(\vec{r}_i))
\end{aligned}
\end{equation}
where the Hamiltonian and $\alpha_{\mathrm{IK}}$ are defined in Section IIB1 of the main text. The compressibility, $\Lambda=\nabla_\Gamma\cdot\dot{\Gamma}$, is computed repeating the steps in Eqs.~\eqref{eq:BeginComp}-\eqref{eq:CompLiouv} of the previous section for the non-equilibrium phase-space flux and is given by
\begin{equation*}
\nabla_{\Gamma}\cdot\dot{\Gamma} = -\beta K \alpha_{\mathrm{IK}}
\end{equation*}

As for the term $\nabla_{\Gamma} f_0 \cdot \dot{\Gamma}$, the gradient of the initial-state distribution is the same of the equilibrium case (see previous section). 
\begin{equation}
\label{eq:GradEqDens}
\nabla_\Gamma f_0 = -\beta f_0 \nabla_\Gamma H_0 + \mathcal{Z}^{-1}\exp[-\beta H_0] \delta'(K - K_0)\nabla_{\Gamma}K
\end{equation} 
The phase space flux of the dynamical system, on the other hand, now corresponds to Eq.~\eqref{eq:NonEqHamEq}.
Multiplication of the first term of Eq.~\eqref{eq:GradEqDens} by the phase-space flux yields
\begin{equation*}
\begin{aligned}
&-\beta f_0\nabla_\Gamma H_0 \cdot \dot{\Gamma} = \beta f_0 \sum_{i=1}^N \vec{\nabla}_{\vec{p}_i}H_0 \cdot \Bigl(-q_i\vec{E} + \frac{\alpha_{\mathrm{IK}}}{2}(\vec{p}_i - q_i\vec{A}(\vec{r}_i))\Bigr) = \\
&= \beta f_0\sum_{i=1}^N \frac{\vec{p}_i - q_i\vec{A}(\vec{r}_i)}{m_i} \cdot \Bigl(-q_i\vec{E} + \frac{\alpha_{\mathrm{IK}}}{2}(\vec{p}_i - q_i\vec{A}(\vec{r}_i))\Bigr) = \\
&= \beta f_0 \Bigl( K \alpha_{\mathrm{IK}} - \sum_{i=1}^N \frac{\vec{p}_i - q_i\vec{A}(\vec{r}_i)}{m_i} \cdot q_i\vec{E} \Bigr)
\end{aligned}
\end{equation*}
The scalar product of the second term of Eq.~\eqref{eq:GradEqDens} with the phase space flux yields again a null contribution as it can be easily shown retracing the steps of the previous section (Eqs.~\eqref{eq:InitNull}-\eqref{eq:FinNull}).

The expression for the dissipation function for the isokinetic system in the chosen external magnetic and electric field is then given by
\begin{equation*}
\begin{aligned}
\Omega^{(0)}(\Gamma) &= -\frac{1}{f_0}\nabla_\Gamma f_0\cdot\dot{\Gamma} - \nabla_\Gamma\cdot\dot{\Gamma} = \\
&= - \beta \biggl( K \alpha_{\mathrm{IK}} - \sum_{i=1}^N q_i\vec{E} \cdot \frac{\vec{p}_i - q_i\vec{A}(\vec{r}_i)}{m_i} \biggr) + \beta K \alpha_{\mathrm{IK}} = \\
& = \beta \sum_{i=1}^N q_i\vec{E} \cdot \frac{\vec{p}_i - q_i\vec{A}(\vec{r}_i)}{m_i}
\end{aligned}
\end{equation*}

\section{Non-interacting particles in external magnetic and electric fields}
\begin{figure}[htbp]
\centering
\includegraphics[width=\columnwidth]{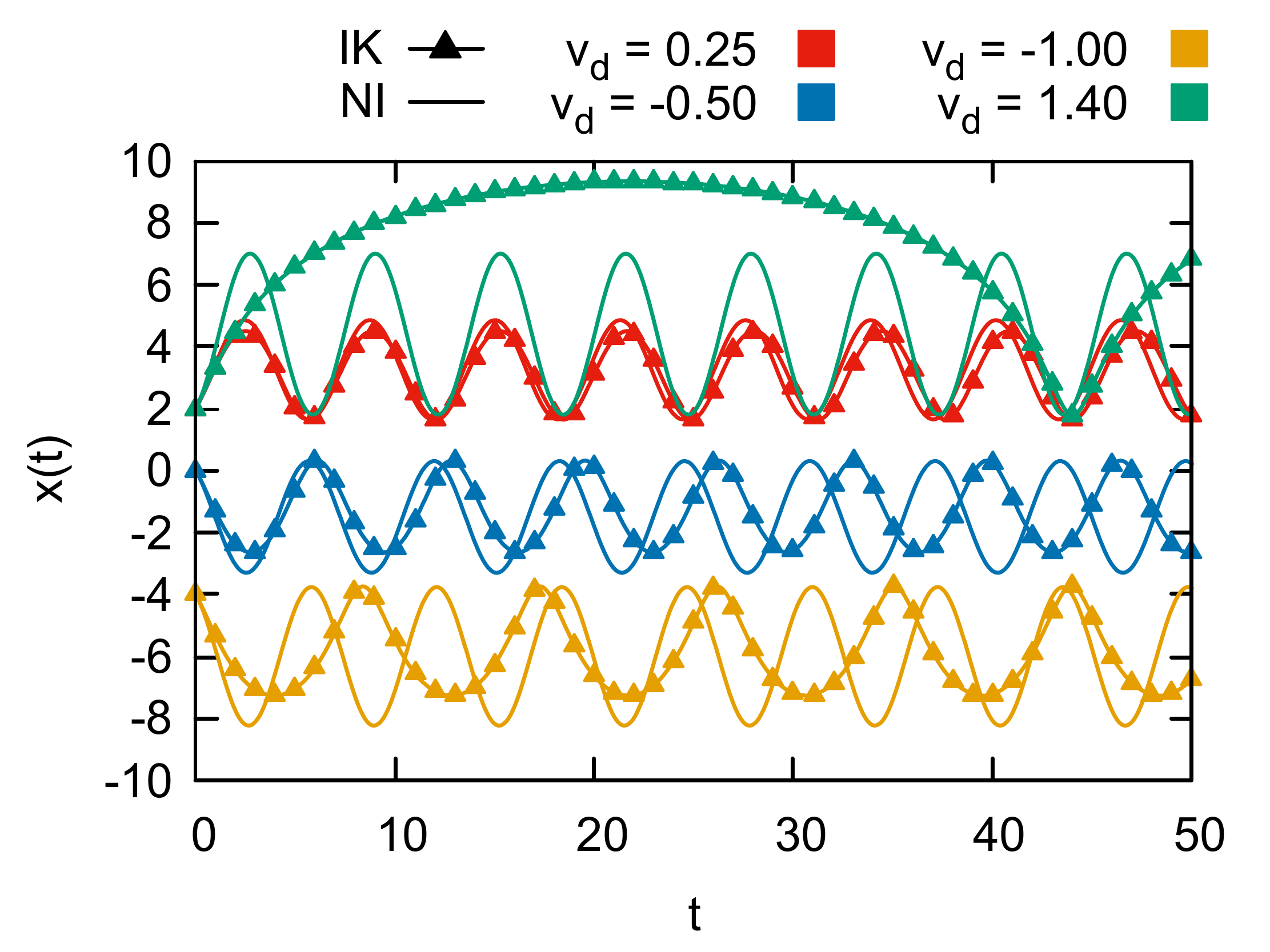}
\caption{Single-particle motion parallel to the electric field for non-interacting (NI, solid curves) and isokinetic (IK, curves with triangles). Color coding refers to evolution for different $v_d$ in the physical regime, with NI and IK solution with the same initial conditions and same drift velocity plotted in the same color. We set $\omega = m = 1$ in arbitrary units. Different initial conditions (leading to $K^* = 1$) are used for different runs and the qualitative behaviour of the trajectory does not change with different (physical) initial conditions. For the thermostatted systems, the physical regime corresponds to $|v_d| < \sqrt{2}$. Non-interacting solutions are analytical, isokinetic evolution is computed numerically using Mathematica~\cite{mathematica:12.1-software}.}
\label{fig:drift}
\end{figure}

Here we discuss the function ${\cal C}_0(A,\delta)$, Eq.~(9) of the main text, for a system of non-interacting particles in external magnetic and electric fields with an isokinetic thermostat. To set the stage, we start by studying the evolution of the system in the absence of the thermostat, an exactly solvable model. The equations of motion of $N$ non-interacting particles in external magnetic and electric field, $\vec{B} = (0,0,B_z)$ and $\vec{E} = (E_x,0,0)$ are given by
\begin{equation}
\begin{aligned}\label{eq:non-interacting}
m_i\ddot{x}_i &= q_iE_x + q_iB_z\dot{y}_i \\
m_i\ddot{y}_i &= -q_iB_z\dot{x}_i \\
m_i\ddot{z}_i &= 0 \\
\end{aligned}
\end{equation}
The system is separable in its single components and the motion on the $z$ axis is trivial. Furthermore, the single-particle solution of Eq.~(\ref{eq:non-interacting}) on the $x$-$y$ plane is~\cite{bittencourt:2004-book}
\begin{equation}
\begin{aligned}
\label{eq:free_solution}
x(t) &= x(0) + \frac{v^\perp}{\omega}\cos(\phi) -\frac{v^\perp}{\omega}\cos(\omega t + \phi) \\
y(t) &= y(0) - \frac{v^\perp}{\omega}\sin(\phi) +\frac{v^\perp}{\omega}\sin(\omega t + \phi) - v_dt
\end{aligned}
\end{equation}
where we have dropped the subscript $i$ for notational convenience and introduced the cyclotron frequency $\omega= \frac{qB_z}{m}$ and the drift velocity $v_d= \frac{E_x}{B_z}$.The constants $v^\perp$ and $\phi$ are fixed by the initial conditions 
\begin{equation}
\begin{aligned}
v^\perp &= \sqrt{\bigl(v^x(0)\bigr)^2 + \bigl(v^y(0) + v_d\bigr)^2} \\
\phi &= \arctan\biggl[\frac{v^x(0)}{v^y(0) + v_d}\biggr]
\end{aligned}
\end{equation}
where $v^{x(y)}(0)$ is the $x$ ($y$) component of the initial velocity. This trajectory is a cycloid on the $x$-$y$ plane
For the non-interacting system, the correlation term in Eq.~(9) of the main text can also be computed analytically. First note that the (single-particle) instantaneous dissipation is given by $\Omegazero(t) = \beta q E_x\dot{x}(t)$. Then
\begin{equation}
\begin{aligned}
\Omega^{(0)}_{0,t} &= \int_0^{t}\dd t \Omegazero(t) = \beta E_x q\int_0^{t}\dd t \dot{x}(t) = \beta E_x q[x(t) - x(0)]\\
\Omega^{(0)}_{t+\tau,2t+\tau} &= \int_{t+\tau}^{2t+\tau}\dd t \Omegazero(t) = \beta E_x q\int_{t+\tau}^{2t+\tau}\dd t \dot{x}(t) = \beta E_x q[x(2t+\tau) - x(t+\tau)]
\end{aligned}
\end{equation}
Substitution of Eqs.~\eqref{eq:free_solution} and some trigonometry yield
\begin{equation}
\begin{aligned}
-\Omega^{(0)}_{0,t}& - \Omega^{(0)}_{t+\tau,2t+\tau} = \\
&-\frac{\beta E_x qv^\perp}{\omega'}\biggl\{\Upsilon(t) + 2\sin\biggl(\frac{1}{2}\omega t\biggr)\biggl[\Theta(t)\cos(\omega\tau) + \Xi(t)\sin(\omega\tau)\biggr]\biggr\}
\end{aligned}
\end{equation}
with
\begin{equation}
\begin{aligned}
\Upsilon(t) &= \cos(\phi) [1-\cos(\omega t)] + \sin(\omega t)\sin(\phi) \\
\Theta(t) &= \cos(\phi)\sin\biggl(\frac{3}{2}\omega t\biggr)+\sin(\phi)\cos\biggl(\frac{3}{2}\omega t\biggr) \\
\Xi(t) &= \cos(\phi)\cos\biggl(\frac{3}{2}\omega t\biggr) - \sin(\phi)\sin\biggl(\frac{3}{2}\omega t\biggr)
\end{aligned}
\end{equation}
The integrated $N$-particle dissipation functions for the separable system, Eq.~(\ref{eq:non-interacting}), are given by the sum of the single-particle quantities computed above. Substituting in Eq.(9) in the main texts results in 
\begin{equation}
\begin{aligned}
{\cal C}_0(A,\delta) &= \lim_{\tau\to\infty} \frac{1}{\tau} \lim_{t\to\infty} \left\langle \exp\biggl[-\beta\sum_i^N\frac{E_xq_iv_i^\perp}{\omega_i}\Bigl\{\Upsilon_i(t)\right. +\\
&+\left.2\sin\Bigl(\frac{\omega_i t}{2}\Bigr)\Bigl[\Theta_i(t)\cos(\omega_i\tau) + \Xi_i(t)\sin(\omega_i\tau)\Bigr]\Bigr\}\biggr]\right\rangle^{(0)}_{\{\overline{\Omegazero}_{t,t+\tau}\}_{(A)_{\delta}}}
\end{aligned}
\end{equation}
The exponent of the expression above remains bounded for all values of $t$ and $\tau$, ensuring that --- as discussed in the main text --- the function ${\cal C}_0(A,\delta)$ is null. This result remains qualitatively true also for the thermostatted system. The evolution of the isokinetic non-interacting model is still separable and trivial along the $z$ axis, with the single-particle evolution on the $x$-$y$ plane (see Eq.~(12) of the main text) given by
\begin{equation}
\label{eq:diff_thermo}
\begin{aligned}
\ddot{x} &= \omega(v_d + \dot{y}) - \frac{m\omega v_d}{2K^*}\dot{x}^2 \\
\ddot{y} &= -\omega\dot{x} - \frac{m\omega v_d}{2K^*}\dot{x}\dot{y}\\
\end{aligned}
\end{equation}
where (see Eq.~(13) in the main text)
\begin{equation}
\label{eq:isokinetic_param}
\alpha_{\mathrm{IK}} = \frac{m\omega v_d}{K^*}\dot{x}
\end{equation}
with $K^* = \frac{1}{2m}\Bigl[\bigl(v^x(0)\bigr)^2+\bigl(v^y(0)\bigr)^2\Bigr]$. 
The system~(\ref{eq:diff_thermo}) cannot be solved analytically, but its properties can be determined combining analysis of its equilibrium solutions, $\ddot{x} = \ddot{y} = 0$, with numerical integration. The study of the equilibrium solutions shows that for $|v_d| > \sqrt{2K^*/m}$, the thermostat dominates the motion leading to a nonphysical constant-velocity diffusion in the plane. Figure~\ref{fig:drift} shows the $x$-component of the trajectory numerically obtained for different drift velocities in the interval $-\sqrt{2K^*/m} < v_d < \sqrt{2K^*/m}$ (the $y$-component does not enter in the evaluation of the single-particle dissipation). In the figure, the numerical results are compared with non-thermostatted evolution for each value of the drift velocity. As it can be seen, the motion remains bounded along the $x$ direction also in the presence of the thermostat, implying --- as for the non-interacting case --- decay of ${\cal C}_0(A,\delta)$. 

The analysis discussed above holds in general for the components of the electric field orthogonal to the magnetic field. If the fields have a parallel component, the magnetic field, which only influences the orthogonal motion, does not affect directly dissipation in the parallel direction.

\end{document}